\documentclass[%
 reprint,
 amsmath,amssymb,
 aps,
]{revtex4-1}

\usepackage{graphicx,subfigure}
\usepackage{dcolumn}
\usepackage{bm}

\begin{document}

\preprint{APS/123-QED}

\title{Correlation between thermodynamic anomalies and pathways of ice nucleation in supercooled water}

\author{Rakesh S. Singh}
\author{Biman Bagchi}
\email{Corresponding author: bbagchi@sscu.iisc.ernet.in}
\affiliation{ Solid State and Structural Chemistry Unit, Indian Institute of Science, Bangalore 560012, India}%

\date{\today}

\begin{abstract}
The well-known classical nucleation theory (CNT) for the free energy barrier towards formation of a nucleus of critical 
size of the new stable phase within the parent metastable phase fails to take into account the influence 
of other metastable phases having density/order intermediate between the parent metastable phase and the final stable phase. 
This lacuna can be more serious than capillary approximation or spherical shape assumption made in CNT. This issue is 
particularly significant in ice nucleation because liquid water shows rich phase diagram consisting of two (high and low 
density) liquid phases in supercooled state. The explanations of thermodynamic and dynamic anomalies of 
supercooled water often invoke the possible influence of a liquid-liquid transition between two metastable liquid phases. 
To investigate both the role of thermodynamic anomalies and presence of distinct metastable liquid phases in supercooled water 
on ice nucleation, we employ density functional theoretical approach to find nucleation free energy barrier in different 
regions of phase diagram. The theory makes a number of striking predictions, such as a dramatic lowering of nucleation barrier 
due to presence of a metastable intermediate phase and crossover in the dependence of free energy barrier on temperature near 
liquid-liquid critical point. These predictions can be tested by computer simulations as well as by controlled experiments.
\end{abstract}

\maketitle

\section{Introduction}
Liquid water shows pronounced thermodynamic (isothermal compressibility, isobaric heat capacity, negative thermal 
expansion coefficient below $4^{o} C$ etc.) as well as dynamic anomalies in its supercooled state~\cite{1, phystoday, 2,3,4, 5, 
6, sengers, shinji, 8}. One popular, yet controversial, 
interpretation of the observed anomalies invokes the concept of liquid-liquid transition and Widom line 
in supercooled water~\cite{9, 35, 36}. Poole \textit{et al.}~\cite{9} 
observed the existence of two liquid phases -- high density liquid (HDL) and low density liquid (LDL) in supercooled 
ST2 model of water. These two liquid phases undergo first order liquid-liquid phase transition (LLPT) on supercooling. 
The HDL-LDL coexistence line ends at liquid-liquid critical point (LLCP). We must note that in simulations one 
observes LLPT at thermodynamic conditions \textit{far from ambient conditions}~\cite{9}. Several recent experiments also suggest the 
existence of liquid-liquid transition in water~\cite{10,11, pnas_exp} as well as in other molecular liquids~\cite{12, 13, 14}. 
However, the unambiguous confirmation of existence of LLPT in bulk supercooled water is hampered by fast ice nucleation at 
thermodynamic conditions predicted for the existence of LLPT. Alternative interpretations of the increase in response functions 
upon supercooling also exist that does not invoke the concept of LLPT~\cite{15,16}.\\
Despite water being the most ubiquitous as well as the most studied liquid on earth, the origin of anomalies in supercooled 
water is still debated~\cite{17, 18, poole_1, 20, 19, 21}. Most surprisingly, while much discussion has focused on the possible presence 
(or absence) of HDL/LDL phases and LLCP~\cite{17, 18, poole_1, 20, 19, 21}, relatively fewer studies have focused on crystallization of 
ice and the effects of pronounced thermodynamic and dynamic anomalies observed in supercooled water on the pathway of ice 
nucleation~\cite{17, 18, 22, 23}.\\
Nucleation of a new phase in correlated molecular systems (such as water) can be a highly complex process. There are 
a multitude of factors (both thermodynamic and dynamic) for this difficulty. The interplay between different length and 
energy scales present in molecular systems gives rise to unusually slow relaxation of the system and also rich phase diagram 
(polymorphism)~\cite{24} in supercooled state. Rich phase behavior can offer diverse (non-classical) nucleation 
pathways~\cite{25, 26, 27}. Both these factors along with observed thermodynamic and dynamic anomalies in supercooled water 
enhance the complexity of ice nucleation from supercooled water and poses fundamental limitations to the classical nucleation 
theory (CNT)~\cite{nuc_1, nuc_2, 28}.
\begin{figure*}[ht!]
     \begin{center}
        \subfigure{
            \label{fig1a}
            \includegraphics[width=0.42\textwidth]{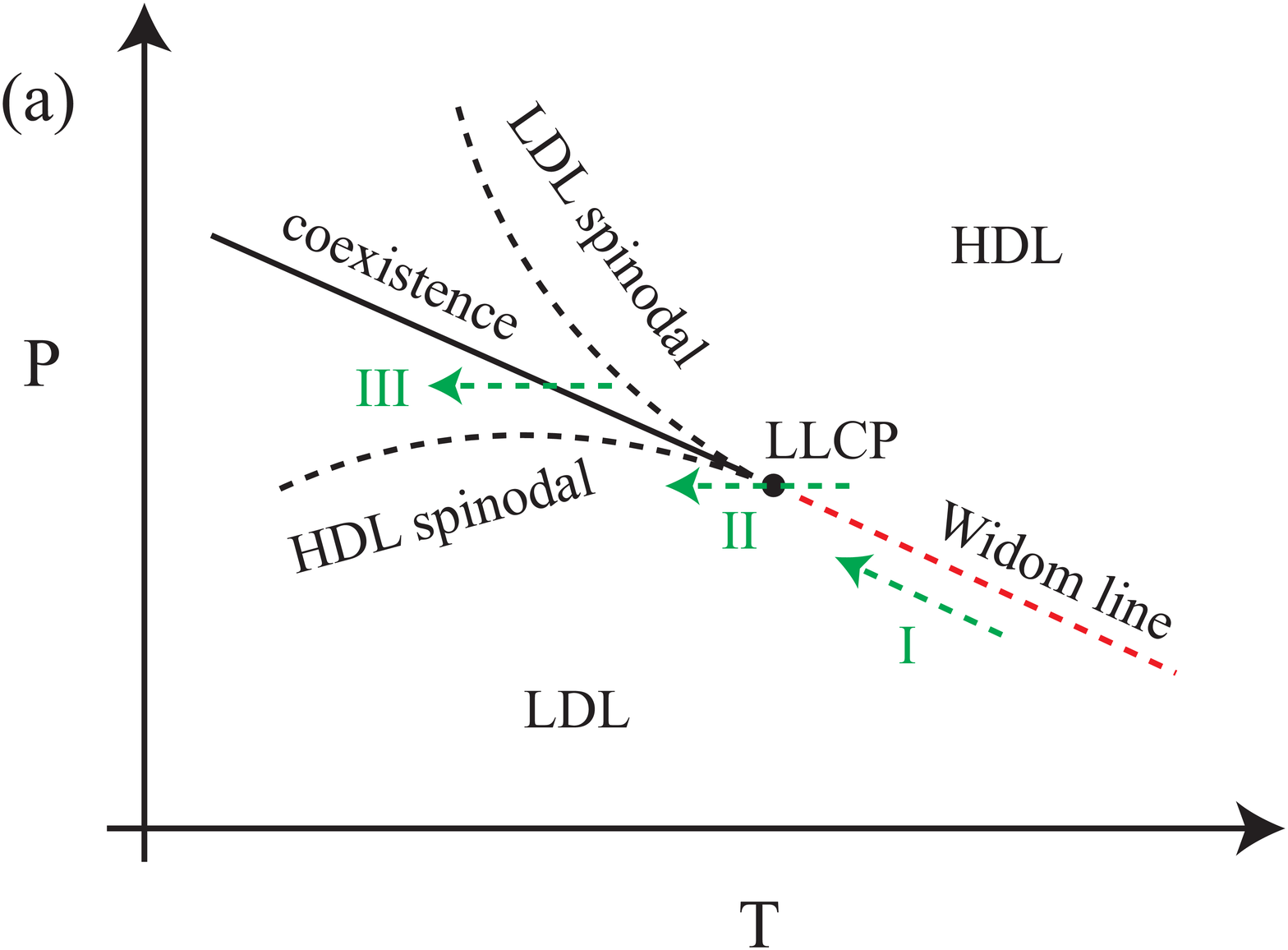}
        }
        \subfigure{
           \label{fig1b}
           \includegraphics[width=0.43\textwidth]{fig_1b.eps}
        }\\ 
        \vspace{10pt}
        \subfigure{
            \label{fig1c}
            \includegraphics[width=0.42\textwidth]{fig_1c.eps}
        }
        \subfigure{
            \label{fig1d}
            \includegraphics[width=0.42\textwidth]{fig_1d.eps}
        }
    \end{center}
    \caption{(a) A schematic phase diagram of supercooled water. LLCP stands for liquid-liquid critical point. 
(b) The free energy surfaces along the green dotted lines indicated by I, II and III are shown in (b), (c) and (d), respectively. 
Note that for simplicity we have only shown the dependence of the curvature of supercooled liquid phase free energy basin on 
changing thermodynamic conditions.}
   \label{fig1}
\end{figure*}
\begin{figure}[ht!]
\vspace{0pt}
     \begin{center}
        \subfigure{
            \label{fig2a}
            \includegraphics[width=0.43\textwidth]{fig_2a.eps}
        }\\
\vspace{0.6cm}
        \subfigure{
           \label{fig2b}
           \includegraphics[width=0.43\textwidth]{fig_2b.eps}
        }\\ 
\vspace{0.6cm}
        \subfigure{
           \label{fig2c}
           \includegraphics[width=0.44\textwidth]{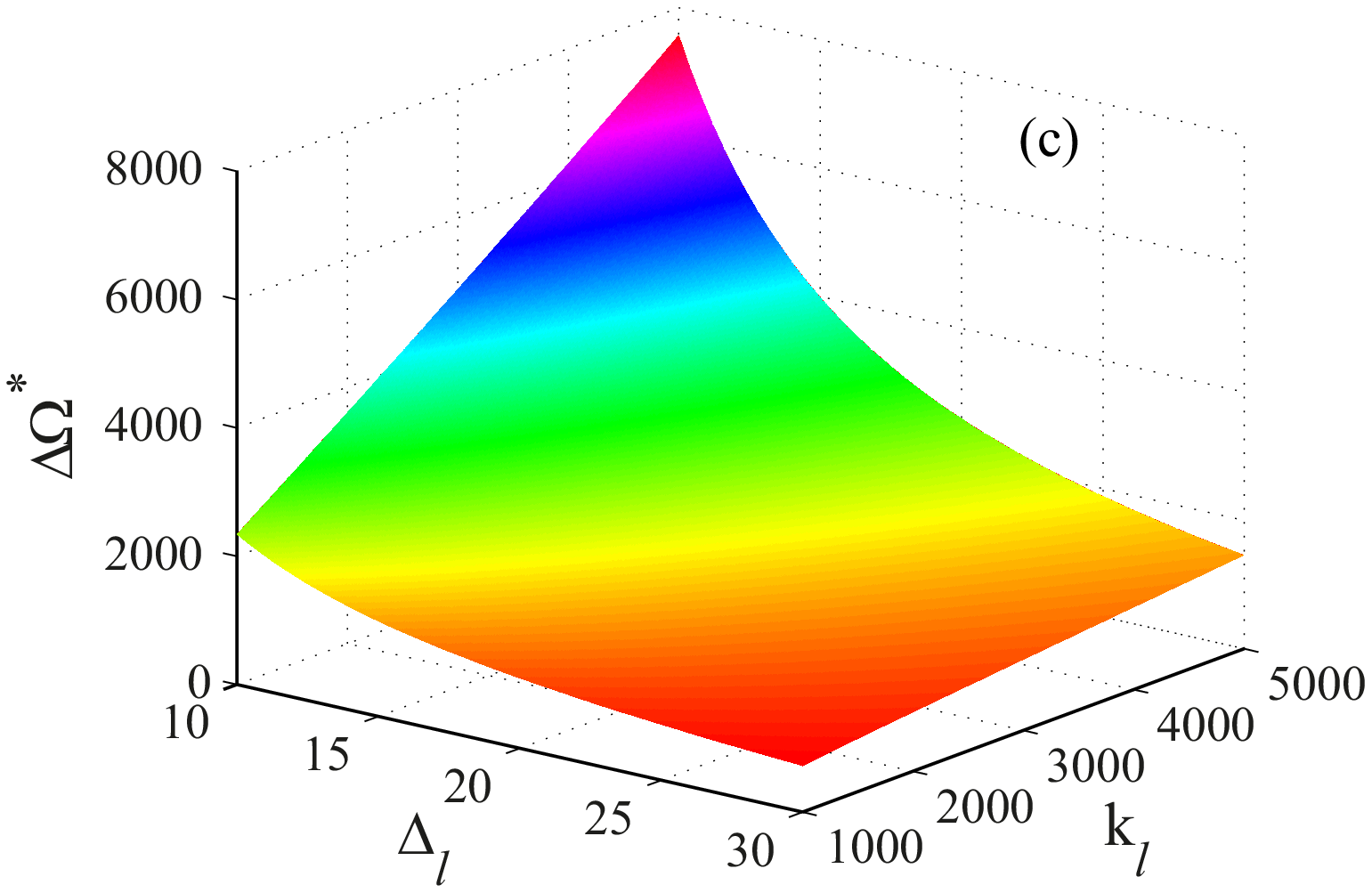}
        }\\ 
    \end{center}
    \caption{(a) Density profiles of the critical cluster for different values of the curvature of liquid free energy basin 
($k_{l}$). We have chosen $k_{i} = 2\times 10^{5}$, density of ice, $\rho_{i} = 0.90$, density of liquid, $\rho_{l} = 1.0$ and 
supercooling parameter, $\Delta_{l} = 25.0$. (b) The dependence of the free energy barrier on $k_{l}$. The dotted line 
(with diamonds) indicates the nucleation barrier predicted by CNT. (c) The dependence of free energy barrier on both curvature 
of liquid free energy basin ($k_{l}$) and free energy gap between supercooled liquid and ice basins ($\Delta_{l}$). Note the 
linear decrease of the free energy barrier as well as flattening of the interface on decreasing $k_{l}$.}
   \label{fig2}
\end{figure}
\section{Density functional theoretical formulation of ice nucleation}
Two major assumptions of CNT~\cite{nuc_1, nuc_2, 28} are (i) the capillary approximation that alllows us to write the free 
energy of a growing nucleus in terms of a sum of the surface energy and the bulk free energy terms, and (ii) the spherical 
shape of the nucleus. It is virtually impossible to remove these assumptions in any self-consistent way without making use of 
substantial numerical work which destroys the simplicity of CNT. However, these approximations might not introduce errors of 
many orders unless we approach the spinodal line.

Density functional theory (DFT)~\cite{29, 30, 31, 32, 33} allows us to address the problem of the free energy of growing 
nucleus without making the capillary approximation. In DFT one gets directly the (unstable) equilibrium 
density profile by minimizing the grand potential of the inhomogeneous system,
\begin{equation} \label{eq1} 
\Omega[\rho(\mathbf{r})] = \int d\mathbf{r}\left[\omega \left(\rho(\mathbf{r})\right) + \frac{1}{2}k\left(\nabla\rho(\mathbf{r})\right)^{2}\right],      
\end{equation} 
with respect to density profile $\left(\rho (\mathbf{r})\right)$ and corresponding free energy cost for formation of the density
 profile of critical cluster. In the above equation, $k$ is density correlation length and $\omega$ is the grand 
potential density. The non-local effects in the system due to inhomogeneity in the density are accounted for in the square 
gradient term. 
\subsection{Phase diagram of supercooled water}
As mentioned, ice nucleation in water could be substantially different due to the possible presence of multiple intermediate 
thermodynamic phases and also from the fact that the participating free energy surfaces (FES's) of the parent and 
daughter phases can be quite different from each other. Thus, the FES of supercooled water can become soft (as we approach $
\approx 230K$) due to multiple reasons. In Fig.~\ref{fig1a}, we have shown a schematic phase diagram of 
supercooled water. Liquid-liquid coexistence line ends at LLCP. Beyond LLCP, the red dotted line shows the Widom line 
(locus of the maxima of thermodynamic response functions). As evident from the figure, depending on the change of controlled 
thermodynamic variables nucleation scenario will differ. The FES's (emphasizing the change in the curvature of liquid free 
energy basin) on changing thermodynamic variables along the dotted green arrows indicated by I, II and III in 
Fig.~\ref{fig1a} are shown in Fig.~\ref{fig1b}, Fig.~\ref{fig1c} and Fig.~\ref{fig1d}, respectively. 

In this article, we have used one order parameter DFT approach to explore the effects of thermodynamic anomalies and distinct 
metastable liquid phases in supercooled water on ice nucleation by studying ice nucleation in different regions of phase 
diagram and made a number of striking predictions that can be tested either by computer simulation or controlled experiments. 
The present analysis offers an explanation of the proximity of homogeneous nucleation temperature to the apparent LLCP. This 
DFT based approach can also be extended to multiple order parameters~\cite{31, 39}.
\subsection{Effects of softening of free energy surface on ice nucleation}
The anomalous increase in thermodynamic response functions on approaching towards LLCP (dotted green arrow I) indicates the 
softening of the metastable liquid FES. As illustrated in Fig.~\ref{fig1a}, on moving along the Widom line towards the LLCP, 
two thermodynamic variables (pressure and temperature) change. Increase in pressure has an opposite effect (favors high density 
phase) to that of decrease in temperature (favors ice-like low density phase) on density as well as free energy gap between 
supercooled water and ice. Considering these opposite dependencies and sake of simplicity we have neglected the thermodynamic 
condition dependence of both the density of supercooled water as well as the relative free energy gap between two phases. 

We include the softening of in FES through Landau type free energy expansions in order parameter(s). The grand potential 
densities of ice and metastable liquid phases are assumed as,   
\begin{equation} \label{eq2} 
\begin{split}
\omega_{i}(\rho) &= \frac{1}{2}k_{i}\left(\rho - \rho_{i}\right)^{2} \\
\omega_{l}(\rho) &= \frac{1}{2}k_{l}(P,T)\left(\rho - \rho _{l}(P,T)\right)^{2} + \Delta_{l}(P,T),
\end{split}
\end{equation} 
where $k_{i}$ is the curvature of ice free energy basin, $\rho_{i}$ is the equilibrium density of ice phase and $\rho_{l}$ is 
the equilibrium density of metastable liquid phase. $\Delta_{l}$ is the supercooling parameter. The values of these 
parameters are mentioned in the caption of Fig.~2.   
At a particular supersaturation we can evaluate the density profile of the critical nucleus by solving analytically the 
Euler-Lagrange equation, $\delta\Omega\left[\rho(\mathbf{r})\right]/\delta\rho(\mathbf{r}) = 0$, 
with appropriate boundary conditions (see Appendix~\ref{a} for details). $\Omega\left[\rho(\mathbf{r})\right]$ is defined by 
Eq.~\ref{eq1} with correlation length, $k = 1\times 10^{4}$ and $\omega\left(\rho\right) = \min\left[\omega_{i}\left(\rho\right)
,\omega_{l}\left(\rho\right)\right]$. The advantage of such a simplified, yet representative, approach is that one can solve 
Euler-Lagrange equation analytically (still-non-trivial).

The expression for density profile of critical cluster is, 
\begin{equation}\label{eq3}
 \begin{split}
  \rho &= \rho_{i} + \frac{\left(\rho_{c} - \rho _{i}\right)r_{c}}{r} \\
& \times \left[\frac{\exp\left(r\sqrt{c_{i}}\right) - \exp\left(-r\sqrt{c_{i}}\right)}{\exp\left(r_{c}\sqrt{c_{i}}\right) - 
\exp\left(-r_{c}\sqrt{c_{i}}\right)}\right],\;\;\;\;\;\;\;\;\; \rho < \rho_{c} \\
\rho &= \rho_{l} + \frac{\left(\rho _{c} - \rho _{l}\right)r_{c} }{r}\exp\left(-(r-r_{c})\sqrt{c_{l}}\right),\;\; \rho > \rho_{c}
 \end{split}
\end{equation}
where $c_{i} = k_{i}/k$ and $c_{l} = k_{l}/k$. $\rho_{c}$ is the density of the system at $r = r_c$. 
Relationship between $\rho_{c} $ and $r_c$ can be established by equating the derivatives of density profiles for 
$\rho < \rho_c$ and $\rho > \rho_c$ at $\rho = \rho_c$. This condition is necessary for the smoothness of the 
composite density profile at $r = r_c$. The supercooling parameter ($\Delta_{l}$) is related with $\rho_c$ as, 
$2\Delta_{l} = k_{i}\left(\rho_{c} - \rho _{i}\right)^{2} - k_{l}\left(\rho_{c} - \rho_{l}\right)^{2}$. The density profiles 
given by Eq.~\ref{eq3} for different curvatures of liquid FES are shown in Fig.~\ref{fig2a}. We note the broadening of the 
interface as well as decrease of the critical cluster size on softening of supercooled liquid FES. The nucleation barrier is 
the extra energy cost (with respect to metastable 
homogeneous liquid phase) for the formation of (unstable) equilibrium density profile of critical cluster 
and is given as, $\Delta \Omega ^{*} = \Omega \left(\rho (r)\right) - \Omega \left(\rho _{l} \right)$. \\
The final expression for the nucleation barrier ($\Delta \Omega ^{*}$) can easily be derived using the analytical expression 
of density profile from Eq.~\ref{eq3} and is  
\begin{equation} \label{eq4} 
\Delta \Omega ^{*} = 2\pi kr_{c} \left(1 + r_{c}\sqrt{c_{l}}\right)\left(\rho _{c} - \rho_{l}\right)\left(\rho_{i} - \rho_{l}\right) - 
\frac{4\pi }{3}r_{c} ^{3}\Delta_{l}.   
\end{equation} 
\\ The effect of softening of the metastable liquid phase FES is also reflected in the plot of 
$\Delta \Omega ^{*}$ \textit{vs}. $k_{l} $, shown in Fig.~\ref{fig2b}. Fig.~\ref{fig2c} consists a two 
dimensional plot showing dependence of $\Delta \Omega ^{*}$ on both $\Delta _{l} $ and $k_{l}$. Note the linear 
increase of $\Delta \Omega ^{*} $ on increasing $k_{l} $ as well as stronger dependence of $\Delta \Omega ^{*}$ on $k_{l}$ at 
low supersaturation compared to high supersaturation (see Fig.~\ref{fig2c}). The dotted line in Fig.~\ref{fig2b} indicates 
the nucleation barrier predicted by CNT, $\Delta \Omega ^{*}_{CNT} = 16\pi\gamma^{3}/{3\Delta_{l}^2}$, where $\gamma$ is the 
surface tension at coexistence. The surface tension at coexistence ($\Delta_{l} = 0$) is the extra energy cost per unit area 
for the formation of planner interface and is given by, 
$\gamma = \left(\Omega \left(\rho (z)\right) - \Omega \left(\rho _{l/ice} \right)\right) / A$, 
where $A$ is the surface area of the planner interface. The coexistence density profile can be easily calculated by solving 
the Euler-Lagrange equation, $\delta\Omega\left[\rho(z)\right]/\delta\rho(z) = 0$, with appropriate boundary conditions (see 
Appendix~\ref{b} for details). The expression for surface tension is,
\begin{equation}
\gamma = \frac{\sqrt{k_{i}k}}{2}(\rho_{c}-\rho_{i})(\rho_{l}-\rho_{i}). 
\end{equation}
As, inclusion of the effects of curvature increases the surface tension, DFT prediction of nucleation barrier is larger than 
the nucleation barrier predicted by CNT with coexistence (planner interface) surface tension. \\
\begin{figure}[ht!]
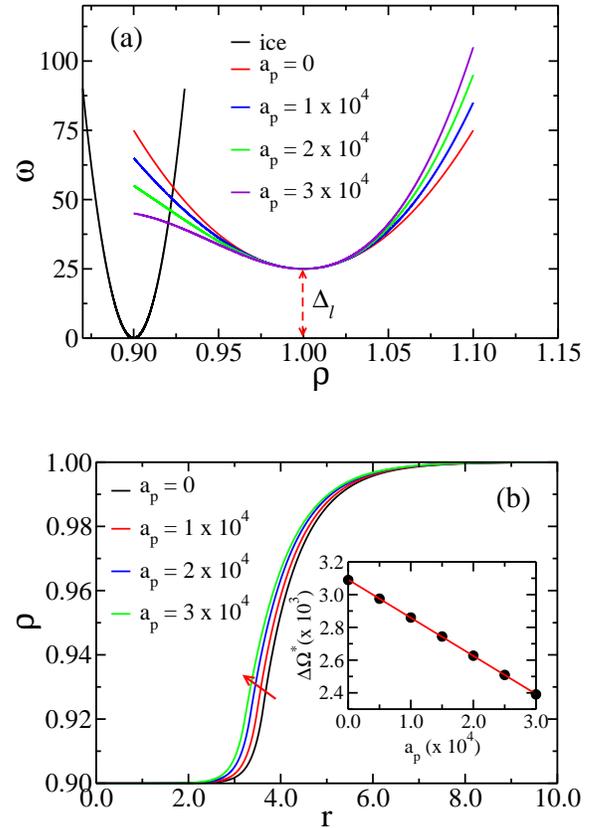

\vspace{10pt}
     \begin{center}
        \subfigure{
            \label{fig3a}
            \includegraphics[width=0.42\textwidth]{fig_3a.eps}
        }\\
\vspace{0.4cm}
        \subfigure{
           \label{fig3b}
           \includegraphics[width=0.42\textwidth]{fig_3b.eps}
        }\\ 
    \end{center}
    \caption{(a) Free energy surfaces as a function of density for ice and metastable liquid phases for different values of 
anharmonicity parameter ($a_{p}$). The chosen parameters are - $k_{i} = 2\times10^{5}$, $k_{l} = 1\times10^{4}$, 
$\rho_{i} = 0.90$, $\rho_{l} = 1.0$ and $\Delta_{l} = 25.0$.(b) Density profiles of critical cluster for different values of 
$a_{p}$. In inset, the dependence of nucleation barrier ($\Delta \Omega ^{*}$) on $a_{p}$ is shown. The points are fitted with 
a straight line, $\Delta \Omega ^{*} = ma_{p} + c$, where $m = -0.0233$  and $c = 3091.78$. Note the linear decrease in 
$\Delta \Omega ^{*}$ on increasing anharmonicity in the liquid phase free energy surface.}
   \label{fig3}
\end{figure}
It should be noted that on increasing supercooling the relaxation time of the 
system also increases and one would expect that the non-equilibrium effects (especially in correlated molecular systems) will 
also become more pronounced. These non-equilibrium effects are not accounted in our study.
\subsection{Beyond the harmonic approximation: role of anharmonicity of free energy surface on nucleation phenomena}
The majority of theoretical studies of phase transitions assume the participating free energy surfaces as harmonic. However, 
in computer simulation studies one often observes an anharmonic softening of the metastable liquid phase FES on supercooling. As 
on increasing anharmonicity, the bulk free energy barrier (which in turn related to surface tension) decreases, one would 
expect that the enhance anharmonic fluctuations on increasing supercooling has pronounced effect on nucleation. In this section, 
we quantify the effects of anharmonic softening of the metastable liquid FES on nucleation barrier and density profile 
of the critical cluster. The grand potential densities of ice and metastable liquid phases are assumed as,
\begin{equation}\label{eq5}
 \begin{split}
\omega_{i}(\rho) &= \frac{1}{2}k_{i}\left(\rho - \rho_{i}\right)^{2} \\
\omega_{l}(\rho) &= \frac{1}{2}k_{l}\left(\rho - \rho_{l}\right)^{2} + a_{p}\left(\rho - \rho_{l}\right)^{3} + \Delta_{l},
\end{split}
\end{equation} 
where $a_{p}$ is the anharmonicity parameter and is a measure of the anharmonicity of the liquid phase FES 
(grand potential density). 

For a fixed supersaturation ($\Delta_{l}$), the grand potential densities for different values of anharmonic parameter are 
shown in Fig.~\ref{fig3a}. The density profile is computed using relaxation method~\cite{34}. In Fig.~\ref{fig3b}, we have 
shown the dependence of the density profile of critical cluster as well as nucleation barrier (see the inset of 
Fig.~\ref{fig3b}) on anharmonicity parameter. Note that on increasing anharmonicity 
of the metastable liquid phase FES, the size of the critical cluster as well as the nucleation free energy barrier 
decreases. This decrease can be attributed to the decrease in the bulk free energy barrier (shown in Fig.~\ref{fig3a}) 
on increasing $a_{p}$. Thus, anharmonic softening of the FES on increasing supercooling has an important role in decreasing 
the free energy barrier of nucleation and hence enhancing the rate of nucleation.
\subsection{Nucleation near liquid-liquid critical point}
\begin{figure}[ht!]
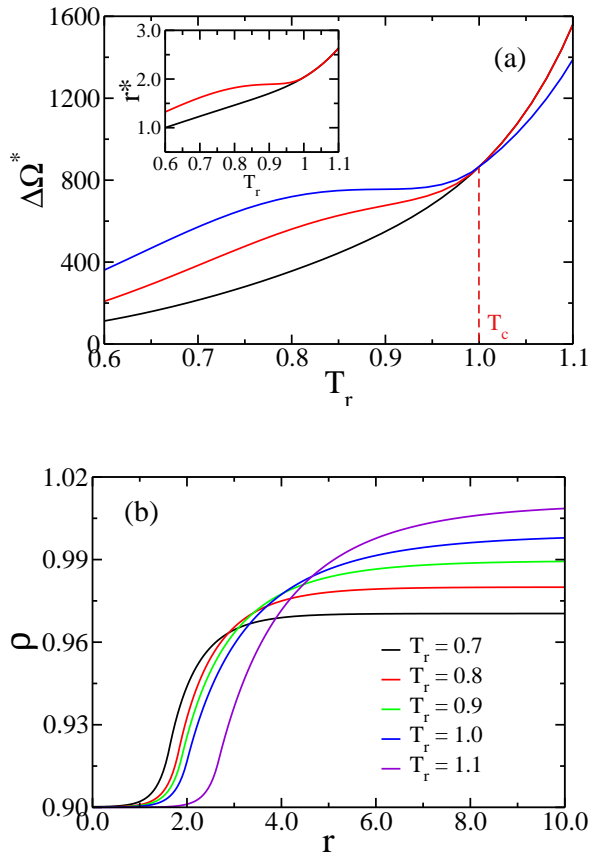

\vspace{10pt}
     \begin{center}
        \subfigure{
            \label{fig4a}
            \includegraphics[width=0.43\textwidth]{fig_4a.eps}
        }\\
\vspace{0.4cm}
        \subfigure{
           \label{fig4b}
           \includegraphics[width=0.43\textwidth]{fig_4b.eps}
        }\\ 
    \end{center}
    \caption{(a) Dependence of nucleation barrier on temperature near liquid-liquid critical point (LLCP). 
The black line indicates the case when curvatures of free energy basins corresponding to LDL and HDL phases are same 
($a_{l} = 1$). The red line indicates the case when the curvature of LDL basin is larger ($a_{l} = 5$) than the HDL basin 
($a_{l} = 1$) and the blue line indicates the case when the free energy gap between the metastable liquid and ice is fixed 
corresponding to its value at $T = T_{c}$ and LDL basin curvature is larger than the HDL basin. In inset, we have shown the 
dependence of critical cluster size on supercooling for the conditions indicated by corresponding colors in  
$\Delta \Omega ^{*}$ \textit{vs.} $T_{r}$ plot. (b) Density profiles of the critical cluster near LLCP at different 
supersaturations for the case when the free energy gap between the metastable liquid and ice is fixed corresponding to its 
value at $T = T_c$ and LDL basin curvature is larger than the HDL basin. Note that temperature is scaled with $T_{c} $.}
   \label{fig4}
\end{figure}

Now, we discuss the nucleation of ice from supercooled water near LLCP (indicated by dotted green arrow II in Fig.~\ref{fig1a}). 
In order to study the effects of metastable LLCP on the ice nucleation scenario we have assumed the grand potential densities 
of ice and supercooled water as,
\begin{equation}\label{eq6}
 \begin{split}
\omega _{i} (\rho ) &= \frac{1}{2} k_{i} \left(\rho - \rho _{i} \right)^{2} \\
\omega _{l} (\rho ) &= \frac{1}{2} k_{l}(T)\left(\rho - \rho _{l}(T)\right)^{2} + \Delta _{l}(T).
\end{split}
\end{equation} 
The temperature dependence of $k_{l}$ is assumed as, $k_{l}(T) = \left(a_{l} \left(T-T_{c} \right)^2 + 1\right)k_{c}$, 
with $T_{c} = 5.0$, $k_{c} = 1000$. $T_{c}$ is the critical temperature and $k_{c}$ is 
the curvature of metastable liquid free energy basin at $T_{c}$. The temperature dependence of density of metastable liquid, 
$\left(\rho _{l}\right)$ is assumed as, $\rho _{l}(T) = \rho_{c} + (T-T_{c})\Delta \rho _{l}$, 
with critical density, $\rho _{c} = 1.0$ and $\Delta \rho _{l} = 0.02$. The grand potential densities corresponding to 
different phases are shown in Fig.~\ref{fig1c}. \\
The calculated nucleation barrier and density profiles are shown in Fig.~\ref{fig4}. The temperature is 
scaled with $T_{c}$, $T_{r} = T/T_{c}$. The blue line indicates the dependence of nucleation 
barrier on temperature when $\Delta _{l} $ is assumed to be independent of temperature ($\Delta_{l} = 25.0$) and $a_{l} = 5$ 
for LDL basin and $a_{l} = 1$ for HDL basin and the black line indicates the same when the curvatures of metastable HDL and 
LDL basins are same ($a_{l} = 1$). The red line indicates the case same as blue line, however, 
the temperature dependence of $\Delta _{l}(T)$  is also taken into account. In this case, $\Delta _{l}$ is varied in such 
as way that $\Delta _{l} = 25.0$ at $T_{r} = 1.0$ starting its values from $\Delta _{l} = 40.0$ at $T_{r} = 0.4$.
As evident from Fig.~\ref{fig4a}, \textit{there is a crossover in the temperature dependence of free energy barrier for ice 
nucleation near LLCP}. Similar crossover behavior is also observed in the temperature dependence of critical cluster size 
(see inset of the figure). We also note that the free energy barrier is relatively insensitive to the supercooling near LLCP. 
\\ In Fig.~\ref{fig4b}, we have shown the temperature dependence of the density profile of critical ice nucleus for the case when 
the free energy gap between the metastable liquid and ice is fixed corresponding to its value at $T = T_{c}$, and LDL basin 
curvature is larger than the HDL basin. Note the increase in the stiffness of density profiles as well as decreases in the 
density difference between bulk phases (ice and supercooled water) on supercooling. Decrease in density difference between 
bulk phases reduces the surface tension cost for formation of ice nucleus; however, at the same time increase in the stiffness 
of density profile (or, curvature of LDL basin) leads to an increase in the surface tension. This delicate balance leads to an 
interesting crossover behavior in the temperature dependence of nucleation barrier near $T_{c}$. This observed crossover 
behavior of ice nucleation barrier can be tested in computer simulation studies. We must note that, depending on the 
stiffness of LDL basin with respect to HDL basin, one might observe a significant increase in the free energy barrier of nucleation 
(in place of being insensitive or weakly sensitive) on increasing supercooling just below $T_c$. The exact nature of crossover 
can only be quantified by inserting more realistic parameters in our theoretical formalism which is not available at the moment.   

The nucleation scenario near LLCP has striking similarity to the (non-classical) pathway of protein crystallization near 
metastable gas-liquid critical point. The metastable critical point enhances density fluctuation and thus 
decreases the nucleation barrier~\cite{37}. One might also observe a similar non-classical nucleation pathway near 
LLCP by invoking a two order parameter description (density and order) where ice nuclei will grow inside low density ice-like 
domains formed due to large scale density fluctuations in the system.
\subsection{Wetting mediated nucleation pathway}
In the last section, we discuss the ice nucleation scenario at thermodynamic conditions where one observes distinct metastable 
minima for LDL and HDL phases (indicated by dotted arrow III in 
Fig.~\ref{fig1a}). The grand potential densities of ice, LDL and HDL phases are assumed as,
\begin{equation} \label{eq7}
\begin{split}
 \omega_{i} (\rho) &= \frac{1}{2}k_{i}\left(\rho - \rho _{i} \right)^{2} \\ 
 \omega_{LDL}(\rho) &= \frac{1}{2} k_{LDL} \left(\rho - \rho_{LDL} \right)^{2} + \Delta_{LDL} \\ 
 \omega_{HDL}(\rho) &= \frac{1}{2} k_{HDL} \left(\rho - \rho_{HDL} \right)^{2} + \Delta_{HDL},
\end{split}
\end{equation}
 \begin{figure}[ht!]
 \vspace{10pt}
     \begin{center}
        \subfigure{
            \label{fig5a}
            \includegraphics[width=0.41\textwidth]{fig_5a.eps}
        }\\
\vspace{0.5cm}
        \subfigure{
           \label{fig5b}
           \includegraphics[width=0.27\textwidth]{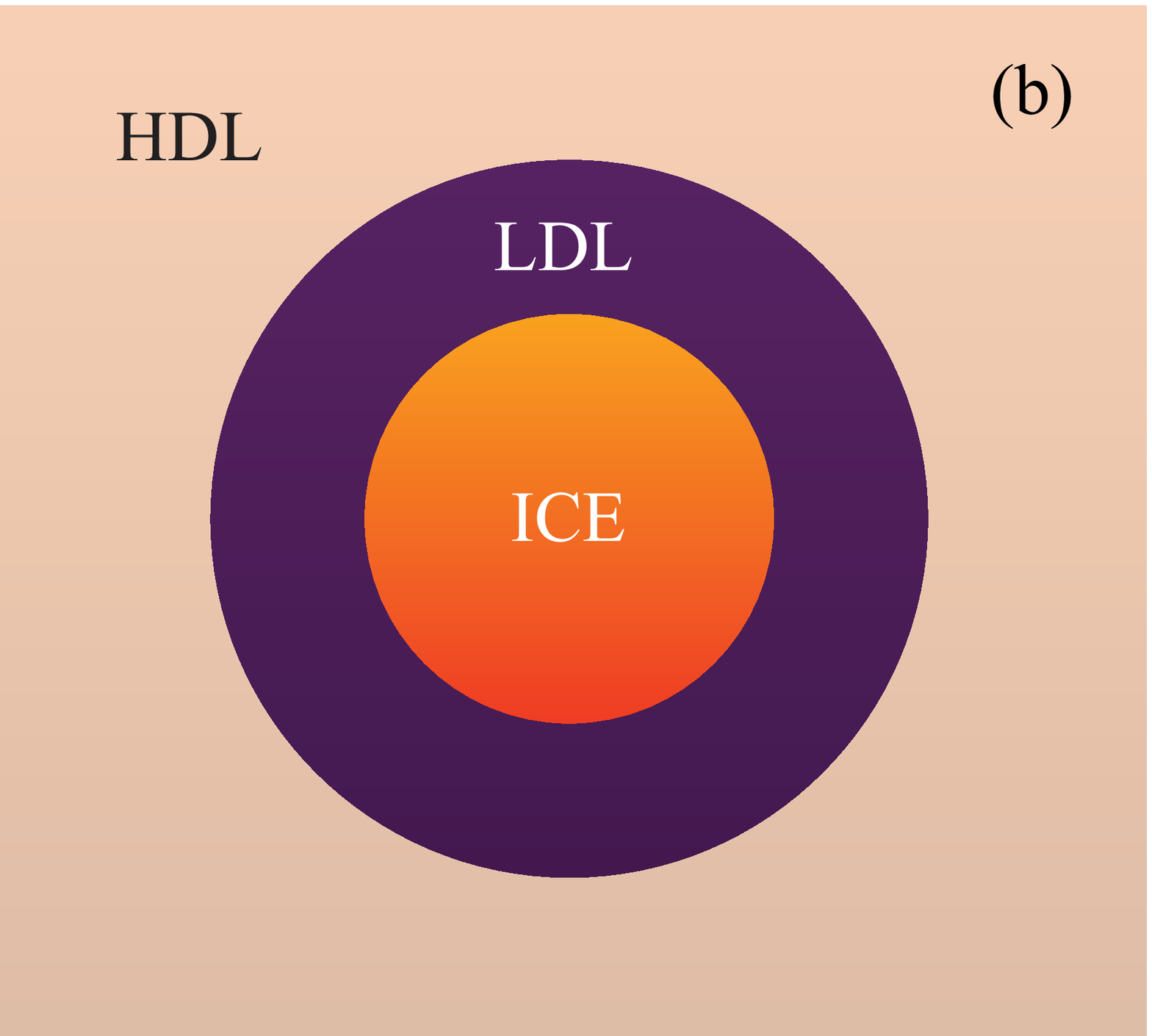}
        }\\ 
    \end{center}
    \caption{(a) The dependence of density profile of critical ice nucleus on the stability of ice phase with respect to HDL 
phase $\Delta_{HDL}$. Note the crossover from wetting mediated transition to one step Ostwald step rule 
type of scenario. For the computation of density profile, we have chosen $\rho_i = 0.90$, $\rho_{LDL} = 0.94$, 
$\rho_{HDL} = 1.05$ and $\Delta_{LDL} = 25.0$. The curvatures of grand potential densities for different phases are $k_i = 2\times 10^5$, 
$k_{LDL} = 5\times10^4$ and $k_{HDL} = 1\times10^4$. (b) A schematic diagram showing the wetting of ice nucleus by an 
intermediate metastable LDL phase.} 
   \label{fig5}
\end{figure}
where $\Delta_{LDL}$ and $\Delta_{HDL}$ are the (meta)stabilities of LDL and HDL phases with respect to ice phase. $\rho_i$, 
$\rho_{LDL}$ and $\rho_{HDL}$ are the densities of bulk ice, LDL and HDL phases, respectively. For numerical computation we 
have assumed the bulk densities and curvatures of grand potential densities of ice, LDL and HDL phases as independent of 
temperature. The grand potential densities of different phases at thermodynamic condition when HDL is metastable with respect 
to both LDL and ice phases are shown in Fig.~\ref{fig1d}. In order to get the density 
profile of critical cluster, we have solved the corresponding Euler-Lagrange equation using relaxation method~\cite{34}.  
The dependence of density profile of critical ice nucleus on (meta)stability of HDL phase with 
respect to ice ($\Delta_{HDL}$) is shown in Fig.~\ref{fig5a}. Note that we have neglected the supercooling dependence of 
$\Delta_{LDL}$ with respect to $\Delta_{HDL}$ ($\Delta_{LDL}$ is fixed at $25.0$), as $\rho_i$ is 
relatively closer to $\rho_{LDL}$ than $\rho_{HDL}$. We have also neglected the non-equilibrium effects arising due to rapid 
increase of relaxation time of LDL phase on decreasing temperature.

As evident from Fig.~\ref{fig5a}, when the HDL phase has minimal metastability with respect to the ice phase, a one step 
density profile (without pronounced wiggle) indicates the absence of (or negligible) wetting of ice nucleus by intermediate 
metastable LDL phase. On gradually increasing the stabilities of ice and LDL phases with respect to HDL phase, we observe a 
significant deviation in the density profile of the critical cluster. This indicates a change in the composition of the 
critical cluster of ice by an intermediate LDL phase. In Fig.~\ref{fig5b}, we have shown a schematic diagram of the 
ice nucleus wetted by an intermediate metastable LDL phase. Note the extent of wetting (width of the metastable LDL region) 
depends on the stability of LDL phase. On further increasing $\Delta_{HDL}$ we observe a transition where a critical cluster 
of intermediate LDL phase appears inside the bulk metastable HDL phase. This is the Ostwald step rule scenario, where transition 
from metastable HDL phase to final stable ice phase occurs via sequential transformations. 

Thus, on increasing supercooling we observe a crossover from the wetting enhanced one step transition to a sequential two 
step (following Ostwald step rule) transition. The existence of a metastable LDL/LDA-like phase, 
with order/density intermediate between HDL and ice can greatly facilitate the nucleation of ice from HDL metastable phase by 
decreasing the surface tension between stable ice and metastable HDL phases. 
Recent computer simulation studies of freezing of water by Matsumoto \textit{et al.}~\cite{22} as well as 
Moore \textit{et al.}~\cite{23} indicate the wetting of ice nucleus by a metastable low density phase. This type of wetting 
mediated transition has also been predicted in the nucleation of liquid from a glassy phase~\cite{38}, 
crystallization of simple (such as hard sphere) and complex (protein crystallization) systems~\cite{37, 38, 39, 40, 41, 42, 43}. The 
application of CNT in the case of wetting mediated transitions is flawed as it does not take into account the decrease 
in surface tension due to wetting (indirect participation of intermediate phase(s)) of the nucleus. 
\section{Conclusion}
To summarize, we have used DFT with phenomenological free energy surfaces to explore the thermodynamic condition dependent 
diverse plausible pathways of ice nucleation from supercooled water. We show that both the softening of the free energy 
surface (due to even a distant presence of 
LLCP in the phase plane) and the presence of distinct metastable liquid phases (LDL and HDL) can lower significantly the 
nucleation free energy barrier. As both the two situations lower the free energy barrier, we need to look 
for the features that can allow us to uniquely identify the actual pathway, at least in computer simulations. The main 
discernible difference occurs in the order parameter profiles as we move from the core of the growing ice nucleus to 
the surface. However, a detailed analysis of the order parameter profile around a critical is yet to be carried out. We also 
observe an interesting crossover in the temperature dependence of nucleation barrier near LLCP.

It should be noted that, in computer simulation studies, the phase diagram as well as formation of ice from supercooled water 
is quite sensitive to the force field used. This often leads to uncertainty and even conflicting results.

\begin{acknowledgments}
We thank Dr. Mantu Santra, Prof. Shinji Saito and Prof. Iwao Ohmine for help and many stimulating discussions. We thank 
the Department of Science and Technology (DST) and the Board of Research in Nuclear Sciences (BRNS), India, for partial 
financial support for this work. B.B. thanks DST for J. C. Bose fellowship.
\end{acknowledgments}
\nocite{*}

\bibliography{jcp}

\begin{appendix}
\section{Analytical expression for nucleation free energy barrier}\label{a}
At a particular supersaturation, density profile of the critical nucleus can be evaluated by solving the 
Euler-Lagrange equation, $\delta\Omega\left[\rho(\mathbf{r})\right]/\delta\rho(\mathbf{r}) = 0$. The resulting equation is 
\begin{equation}
 k\mathbf{\nabla}^2\rho(\mathbf{r}) = \frac{d\omega(\rho(\mathbf{r}))}{d\rho(\mathbf{r})}.
\end{equation}
The above equation in spherical coordinate can be rewritten as,
\begin{equation}
 \frac{d^2\rho}{dr^2} + \frac{2}{r}\frac{d\rho}{dr} = \frac{d\omega}{d\rho}, 
\end{equation}
 where $\omega\left(\rho\right) = \min\left[\omega_{i}\left(\rho\right),\omega_{l}\left(\rho\right)\right]$. This equation as 
close resemblance with free particle time independent Schrödinger equation in spherical coordinate. One can easily derive an 
analytical expansion for density profile by substituting $\rho - \rho_{i/l} = y$ and $y = u/r$. The boundary conditions are, 
$d\rho/dr = 0$ at $r = 0$ and $\rho = \rho_l$ at $r \to \infty$. \\
The expression for density profile is, 
\begin{equation}
 \begin{split}
  \rho &= \rho_{i} + \frac{\left(\rho_{c} - \rho _{i}\right)r_{c}}{r} \\
& \times \left[\frac{\exp\left(r\sqrt{c_{i}}\right) - \exp\left(-r\sqrt{c_{i}}\right)}{\exp\left(r_{c}\sqrt{c_{i}}\right) - 
\exp\left(-r_{c}\sqrt{c_{i}}\right)}\right],\;\;\;\;\;\;\;\;\; \rho < \rho_{c} \\
\rho &= \rho_{l} + \frac{\left(\rho _{c} - \rho _{l}\right)r_{c} }{r}\exp\left(-(r-r_{c})\sqrt{c_{l}}\right),\;\; \rho > \rho_{c}
 \end{split}
\end{equation}
The relationship between $\rho_c$ and $r_c$ can be established by equating, 
\begin{equation}\label{rcrhoc}
 \left.\frac{d\rho}{dr}\right\vert_{r=r_c}^{\rho < \rho_c} = 
\left.\frac{d\rho}{dr}\right\vert_{r=r_c}^{\rho > \rho_c}. 
\end{equation}
For each $\rho_c$, one can get $r_c$ by solving numerically the non-linear algebraic equation given by Eq.~\ref{rcrhoc}.  
The free energy barrier of nucleation is the grand potential for formation of critical cluster in bulk metastable liquid phase,
 \begin{equation}
 \begin{split}
  \Delta \Omega ^{*} &= \Omega \left(\rho (r)\right) - \Omega \left(\rho _{l} \right) \\
   &= 4\pi  \int_{0}^{\infty}drr^2\left[\omega \left(\rho(r)\right) + \frac{1}{2}k\left(\nabla\rho(r)\right)^{2}
   - \omega(\rho_l)\right].
 \end{split}
\end{equation}
The above equation can be splitted as, $\Delta \Omega ^{*} = \Delta \Omega _{i} ^{*} + \Delta \Omega _l ^{*}$, where
\begin{equation}
 \Delta \Omega_i ^{*} = 
    4\pi\int_{0}^{r_c}drr^2\left[\omega_i \left(\rho(r)\right)+ \frac{1}{2}k\left(\nabla\rho(r)\right)^{2} \right.
        -\left.\omega(\rho_l)\vphantom{\frac{1}{2}}\right], 
\end{equation}
and 
\begin{equation}
 \Delta \Omega_l ^{*} = 
    4\pi\int_{r_c}^{\infty}drr^2\left[\omega_l \left(\rho(r)\right)+ \frac{1}{2}k\left(\nabla\rho(r)\right)^{2} \right. 
        -\left.\omega(\rho_l)\vphantom{\frac{1}{2}}\right]. 
\end{equation}
Now, on integrating the above equation and on further simplification using the relationship between $\rho_c$ and $r_c$ 
(Eq.~\ref{rcrhoc}), the expression for nucleation barrier reduces to
 \begin{equation} 
\Delta \Omega ^{*} = 2\pi kr_{c} \left(1 + r_{c}\sqrt{c_{l}}\right)\left(\rho _{c} - \rho_{l}\right)\left(\rho_{i} - \rho_{l}\right) - 
\frac{4\pi }{3}r_{c} ^{3}\Delta_{l}.   
\end{equation} 
\section{Analytical expression for surface tension at coexistence}\label{b}
 Surface tension is the extra energy cost for the formation of planner interface between two coexisting 
phases at equilibrium and is defined as, $\gamma = \left(\Omega \left(\rho (z)\right) - \Omega \left(\rho _{l/ice} 
\right)\right) / A$, where $A$ is area of the planner interface and $\Omega \left(\rho (z)\right)$ is grand potential for 
inhomogeneous density profile $\rho (z)$ and $\Omega \left(\rho _{l/ice}\right)$ is the grand potential of the bulk phases at 
coexistence. Following similar procedure one can derive an expression for surface tension, 
\begin{equation}
\gamma = \frac{\sqrt{k_{i}k}}{2}(\rho_{c}-\rho_{i})(\rho_{l}-\rho_{i}). 
\end{equation}
Now, when the curvature of the grand potential densities corresponding to ice and liquid phases are same, \textit{i.e.} $k_i = 
k_l = k'$, then $\rho_c = (\rho_i + \rho_l) / 2$ and the expression for surface tension reduces to the following well-known 
expression of surface tension initially derived by Cahn and Hilliard~\cite{29},
 \begin{equation}
\gamma = \frac{\sqrt{k'k}}{4}(\Delta \rho)^2, 
\end{equation}
where $\Delta \rho = (\rho_l - \rho_i)$ is the order parameter difference between two coexisting phases and $k$ is the 
correlation length. 
\end{appendix}

\end{document}